\documentclass[12pt]{article}
\usepackage{amssymb}
\usepackage{setspace}
\usepackage{amsmath}

\newcommand{\wt}{\widetilde}
\newcommand{\wh}{\widehat}
\newcommand{\p}{*}
\begin{document}

\title{Multiscaling for Classical Nanosystems: Derivation of Smoluchowski \& Fokker-Planck Equations}
\author{\textsc{S. Pankavich}\footnote{Department of Mathematics, Indiana University, Bloomington, IN
47405, sdp@indiana.edu}, \textsc{Z. Shreif}\footnote{Center for Cell and Virus Theory;
Department of Chemistry, Indiana University, Bloomington, IN 47405, ortoleva@indiana.edu},
\textsc{P. Ortoleva}\footnotemark[2]\\}
\date{\today}
\maketitle

\begin{abstract}
Using multiscale analysis and methods of statistical physics, we show that a solution to the
$N$-atom Liouville Equation can be decomposed via an expansion in terms of a smallness parameter
$\epsilon$, wherein the long scale time behavior depends upon a reduced probability density that
is a function of slow-evolving order parameters.  This reduced probability density is shown to
satisfy the Smoluchowski equation up to $O(\epsilon^2)$ for a given range of initial conditions.
Furthermore, under the additional assumption that the nanoparticle momentum evolves on a slow
time scale, we show that this reduced probability density satisfies a Fokker-Planck equation up
to $O(\epsilon^2)$. This approach has applications to a broad range of problems in the nanosciences.
\end{abstract}

\textbf{Keywords:} nanosystems, all-atom multiscale analysis (AMA), Gibbs Hypothesis,
Smoluchowski equations, Fokker-Planck equations

\section{Introduction}
Nanosystems are currently of great interest in the fundamental and applied life sciences.  A
major unresolved challenge is to develop a predictive approach to these systems that capture the
inter-communication among the processes operating on differing scales in space and time.  The
premise of the present work is that one can introduce order parameters (slowly varying
quantities that capture the essence of large-scale bionanosystem phenomena) and then, using
Newton's equations for the $N$-atom system, derive equations for stochastic order parameter
dynamics.\\

Examples of bionanosystems abound in nature and medicine.  Viruses are supra-million atom
entities with complex structural and functional characteristics, including dramatic transitions,
interactions with host cells, and self-assembly of subunits.  Ribosomes are of size and
complexity similar to viruses, and mediate an important intercellular process - translation of
mRNA into proteins.  Protein nanoviruses conduct electric currents, allowing some bacteria to
exploit oxide mineral grains when performing oxidation in the oxygen-poor subsurface.  In
addition to these natural phenomena, scientists are currently developing nanocapsules for the
delivery of therapeutic payloads (such as drugs, siRNA, or genes) to diseased tissues, and for
medical imaging by equipping nanoparticles with flourescent subunits while other subunits are
designed to bind with diseased cell membrane-bound proteins.  Finally, researchers are designing
mutated viruses with diminished viralence to serve as vaccines.\\

To address all of these applications, one would like to develop predictive modules with the
ability to efficiently simulate the dynamics of bionanosystems.  Such programs should include
the following characteristics :

\begin{itemize}
\item an underlying all-atom description to evaluate the interaction of bionanosystems with selected molecules, membranes, or other features in their background microenvironment\\

\item a model that does not require recalibration with each new application\\

\item an approach that builds in the detailed physical molecular laws and the predictive power following from them\\

\item an approach that is computationally feasible.\\
\end{itemize}

Considering this list, we suggest that a multiscale analysis of the equations of $N$-atom
physics will fulfill each of these requirements.  Molecular Dynamics (MD) is a current
state-of-the-art software package that efficiently performs simulations of Newton's equations
for each of $N$ atoms in a system of interest.  An efficient MD code, NAMD, has previously been
used to simulate a whole virus using a $1024$ CPU supercomputer, but the process proceeds at a
rate of about $1$ nanosecond of simulated time per day.  The typical timescale for a viral
structural transition is on the order of a millisecond or greater.  Thus, the aforementioned MD
code and hardware would take $3,000$ years or longer to attain meaningful results.  As
bionanosystems evolve due to the cross-talk between processes which take place on many scales in
both space and
time, a computational algorithm based on a multiscale approach seems like a natural choice.\\

The use of multiscale techniques in statistical mechanics beginning with the Liouville equation
has a long history (see \cite{SOa}, \cite{Pa}, \cite{Pb}, and \cite{SOb}, and more recently
\cite{Ort}, \cite{MOa}, and \cite{MOb}).  In the present work, we demonstrate several new
elements of the analysis.  First, in our approach, the nanoparticle's internal atomic state, as
well as that of the microenvironment, are maintained allowing for a more natural, symmetric
starting description.  Additionally, we utilize a version of the Gibbs postulated equivalence of
ensemble and long-time averages, following classical results within ergodic theory.  A precise
representation is obtained for the momentum factor in the normalization constant for the
lowest-order $N$-atom probability density in a perturbation expansion of a solution to the
Liouville equation.  As a result, Fokker-Planck and Smoluchowski equations are derived which
describe the stochastic dynamics of these slow-evolving order parameters.  These results can
then be utilized in the production of an efficient software module that can model nanoparticle
behavior over long time scales, thereby capturing the
necessary structural dynamics of a virus.\\

In recent investigations (see \cite{MOa} and \cite{MOb}), the reduced probability density $W$
was shown to obey an unconserved equation of Fokker-Planck type up to $O(\epsilon^2)$. The
derivation of this equation was inconsistent with the mathematical framework of differential
equations as the thermal average and the derivatives with respect to order parameters
$\displaystyle \frac{\partial}{\partial \Phi}$ and $\displaystyle \frac{\partial}{\partial \Pi}$
do not commute. In the work that follows, we eliminate ambiguities regarding the permutation of
the thermal average and these derivatives. Additionally, the lowest order distribution was
previously taken to be independent of the conjugate momentum $\Pi$.  This is done in error,
causing the lowest order dependence on the slow variable $\Pi$ to be lost and propagating this
throughout the multiscale analysis. In Section $3$, we rigorously correct these mistakes and
establish many of the ideas of \cite{MOb} on a more precise footing by showing that the
correction to the reduced probability density $\wt{W}$ indeed satisfies a Fokker-Planck equation
in conservative form. Prior to this, we show in Section $2$ that if the momentum is an
atomically varying quantity, rather than a slowly varying order parameter, then $\wt{W}$
directly satisfies the Smoluchowski diffusion equation up to $O(\epsilon^2)$. In both sections,
our derivations occur from the starting point of the general kinetic equation so that the
resulting coarse-grained equations do not arrive from solubility conditions, but from a rational
expansion of the Liouville equation.\\

\section{Multiscale Analysis: Smoluchowski Equation}

A central goal of multiscale analysis is to rigorously derive coarse-grained equations starting
from a more fundamental, final scale theory. The Liouville equation has been a common starting
point. The challenge is that while the Liouville equation preserves probability by construction,
it is not guaranteed that a given truncation of the equation will be conserving. A re-examination of multiscale analysis for the Liouville equation is now carried out to identify potential difficulties of this type that may arise, and to set forth techniques to resolve them.  In this section we resolve probability conservation violations when the momentum is not a slow variable.\\

Consider the Liouville equation in a multiscale framework wherein order parameters are
introduced.  We consider an $N$-atom system consisting of a nanoparticle of $N^*$ atoms and a
host medium of $N - N^*$ atoms.  For each atom $i = 1,...,N$, we write $p_i$, $r_i \in
\mathbb{R}^3$, and $m_i > 0 $ as the momentum, position, and mass of atom $i$ respectively.  In
addition, we use the notation $\Gamma = \{r_1,p_1,...,r_N,p_N\}$.  For each $i=1,...,N$, define
the indicator function $$
\theta_i = \left \{ \begin{array}{rl} 1, & \mbox{if atom} \ i \ \mbox{is in the nanoparticle} \\
0, & \mbox{otherwise.} \end{array} \right. $$ For the nanoparticle, we define its total mass
$$m = \sum_{i=1}^N m_i \theta_i, $$ the center of mass
\begin{equation}
\label{R} R = \sum_{i=1}^N \frac{m_i}{m} r_i \theta_i,
\end{equation}
and the total momentum
\begin{equation}
\label{P} P = \sum_{i=1}^N p_i \theta_i.
\end{equation}

To begin the multiscale analysis, we first introduce a dimensionless scaling parameter
$\epsilon$ in the mass terms by writing
\begin{equation}
\label{epsilon} \epsilon = \frac{\wh{m}}{m},
\end{equation}
where $\wh{m}$ is the mass of a typical atom. In the case that all atoms in the nanoparticle
have the same mass, $m_i = \hat{m}$ for all $i=1,...,N^\p$, it follows that $\displaystyle
\frac{\wh{m} }{m} = \frac{1}{N^\p}$. Hence, $\epsilon \approx \left (N^\p \right)^{-1}$. In this
section, we make the following assumptions:
\begin{enumerate}
\item The total nanoparticle momentum does not evolve slowly -  $P$ is $O(\epsilon^0)$.
\item The net force on the nanoparticle is not decreased due to cancelation of atomic
contributions - $f$ is $O(\epsilon^0)$.
\item Large migration distances are not a consideration - $R$ is $O(\epsilon^0)$.
\end{enumerate}
As a result, order parameters are $O(\epsilon^0)$ and need not be scaled in $\epsilon$, even
though Newton's equations show that they evolve slowly as $\displaystyle \frac{dR}{dt} =
\frac{P}{m} = \epsilon \frac{P}{\hat{m}} = O(\epsilon)$.  We note that other scalings would be
appropriate to capture different behavioral regimes.\\

Let us assume $\rho$ satisfies the Liouville Equation
\begin{equation}
\label{Liouville} \frac{\partial \rho}{\partial t} = - \sum_{i=1}^N \left [ \frac{p_i}{m_i}
\cdot \frac{\partial}{\partial r_i} + F_i \cdot \frac{\partial}{\partial p_i} \right ] \rho
\equiv \mathcal{L} \rho
\end{equation}
where we define $F_i$ to be the force on atom $i$ and $t$ to be time.  In addition, we assume throughout that $\rho$ decays at infinity (a standard assumption for a probability density) so that boundary terms do not appear in the calculations from integration by parts.  Denote the collection of all atomic positions by $\Gamma_r = \{r_1,...,r_N\}$.  Given the probability
density, $\rho(\Gamma,t)$, we define
\begin{equation}
\label{Wtilde} \wt{W}(R,t) = \int  \Delta(\Gamma^\p_r,R) \rho(\Gamma^\p,t) d\Gamma^\p
\end{equation}
where
\begin{equation}
\label{delta} \Delta(\Gamma^\p_r,R) = \delta(R - R^\p),
\end{equation} $R$ is the center of mass order parameter, and $$R^\p = \sum_{i=1}^N \frac{m_i}{m} r_i^\p \theta_i$$ is the $\Gamma^\p_r$-dependent
value of $R$.  Then, using the dependence of $\wt{W}$ on $\rho$, a solution to the Liouville
equation, we may show that $\wt{W}$ must satisfy a conserved equation.  Since $\rho$ satisfies
(\ref{Liouville}), we find
\begin{eqnarray*}
\frac{\partial \wt{W}}{\partial t} & = & \int  \Delta \frac{\partial \rho}{\partial t}(\Gamma^\p,t) d \Gamma^\p \\
& = & - \int  \Delta \left ( \sum_{i=1}^N \frac{p_i^\p}{m_i} \cdot \frac{\partial
\rho}{\partial r_i^\p} + F_i \cdot \frac{\partial \rho}{\partial
p_i^\p} \right ) d \Gamma^\p \\
& = & \int  \rho \left ( \sum_{i=1}^N \frac{p_i^\p}{m_i} \cdot \frac{\partial
\Delta}{\partial r_i^\p} + F_i \cdot \frac{\partial \Delta}{\partial p_i^\p}
\right ) d \Gamma^\p \\
& = & - \int  \rho \left (\sum_{i=1}^N \frac{p_i^\p}{m_i} \cdot \frac{\partial
R^\p}{\partial r_i^\p} + F_i \cdot \frac{\partial R^\p}{\partial
p^\p_i} \right ) \frac{\partial \Delta}{\partial R} d \Gamma^\p \\
& = & \frac{\partial}{\partial R} \left ( \int  \rho \Delta \mathcal{L}R^\p d \Gamma^\p \right )\\
& = & - \frac{\partial}{\partial R} \left ( \int  \rho \Delta \frac{P^\p}{m} d \Gamma^\p \right
)
\end{eqnarray*}
Thus, the reduced probability density, $\wt{W}$, satisfies
\begin{equation}
\label{consW} \frac{\partial \wt{W}}{\partial t} = - \epsilon \frac{\partial}{\partial R} \left
( \int  \rho \Delta \frac{P^\p}{\wh{m}} d \Gamma^\p \right ).
\end{equation}

Next, we attempt to determine $\rho$ up to $O(\epsilon)$.  The N-atom probability density,
$\rho(\Gamma,t)$, is then assumed to be expressed as a function of an additional argument,
$\Upsilon(\Gamma,t,\cdot)$ in such a way that when the last argument is evaluated at $R$, $\rho$
is obtained, i.e $\rho(\Gamma,t) = \Upsilon(\Gamma,t,R)$.  Instead of labeling this new
function, we will just extend our previous notation and refer to it as $\rho(\Gamma,t,R)$. This
displays the dependence of the probability density on multiple scales of motion.  Hence, $\rho$
depends on the all-atom descriptive variables $\Gamma$, as well as on $R$ defined by (\ref{R}),
the latter an expression of the fact that $\rho$ has indirect dependence on the all-atom state
through order parameters and thus depends on the all-atom state in several, distinct ways.\\

We apply the Liouville operator to $\rho(\Gamma,t,R)$ and invoke the chain rule to find
$$ \frac{\partial \rho}{\partial t} = - \mathcal{L}_0 \rho - \sum_{i=1}^N \frac{p_i}{m_i}
\cdot \frac{dR}{d r_i} \frac{\partial \rho}{\partial R}.$$  Using (\ref{R}) this becomes
\begin{equation}
\label{chain} \frac{\partial \rho}{\partial t} = - \mathcal{L}_0 \rho - \frac{P}{m} \cdot
\frac{\partial \rho}{\partial R}.
\end{equation}
Here, we are writing $\mathcal{L}_0$ instead of $\mathcal{L}$ because these derivatives are
taken at constant values of $R$.  By introducing (\ref{epsilon}) into (\ref{chain}), the
Liouville equation (\ref{Liouville}) transforms into a multiscale equation (see \cite{MOa} and
\cite{MOb} for more details) as

\begin{equation}
\label{Multiscale} \frac{\partial \rho}{\partial t} = \left ( \mathcal{L}_0 + \epsilon
\mathcal{L}_1 \right ) \rho
\end{equation}
where
\begin{equation}
\label{L0} \mathcal{L}_0 = - \sum_{i=1}^N \left [ \frac{p_i}{m_i} \cdot \frac{\partial}{\partial
r_i} + F_i \cdot \frac{\partial}{\partial p_i} \right ]
\end{equation}
and
\begin{equation}
\label{L1} \mathcal{L}_1 = - \frac{P}{\wh{m}} \cdot \frac{\partial}{\partial R}.
\end{equation}
Again, it must be noted that $\mathcal{L}$ and $\mathcal{L}_0$, while seemingly exact in
definition, differ because the differentiation in $\mathcal{L}_0$ is performed at constant
values of order parameters $R$.  Additionally, the differentiation in $\mathcal{L}_1$ is
performed at fixed values of $\Gamma$.  Further details regarding the all-atom, multiscale
analysis (AMA) for the Liouville equation can be found in \cite{MOa}, \cite{MOb}, \cite{Ort},
and \cite{SO}.  The operators (\ref{L0}) and (\ref{L1}) differ from that of the previously
mentioned papers \cite{MOa} and \cite{MOb} since in that work the conjugate momentum $\Pi$ is
treated as an order parameter.  In
the work that follows in this section, we treat this momentum term as a micro-variable instead.\\

Assuming the net force on the nanoparticle does not experience cancelation due to fluctuating
terms, it can be written in terms of the individual atomic forces as
\begin{equation}
\label{f} f = \sum_{i=1}^N F_i \theta_i.
\end{equation}
We let $V(\Gamma_r)$ be the $N$-atom potential so that $\displaystyle \frac{\partial V}{\partial
r_i} = -F_i$ for every $i = 1,...,N$.  Next, we assume that $\rho$ may be expressed as a power
series in $\epsilon$:
\begin{equation}\label{II4}
\rho = \sum_{n=0}^\infty \rho_n \epsilon^n
\end{equation}
A set of time variables, defined via $t_n = \epsilon^n t$, is introduced to capture effects of
processes occurring on the various timescales. The chain rule implies
\begin{equation}\label{II5}
\frac{\partial}{\partial t} = \sum_{n=0}^\infty \epsilon^n \frac{\partial}{\partial t_n}.
\end{equation}
We then expand (\ref{Multiscale}) using (\ref{II4}) and (\ref{II5}) and separate $\epsilon$
scales.  Define for $n \in \{0\} \cup \mathbb{N}$,
\begin{equation}
\label{Lambdan} \Lambda_n = \frac{\partial}{\partial t_n} - \mathcal{L}_n
\end{equation}
where we take $\mathcal{L}_n = 0$ for $n > 1$.  The expansion yields the equations
$$ \Lambda_0 \rho_0 = 0,$$ and for $n \in \mathbb{N}$, $$ \Lambda_0 \rho_n = - \sum_{i=1}^n \Lambda_i
\rho_{n-i}.$$

Assuming the statistical state of the system has quasi-equilibrium character, the lowest order
distribution $\rho_0$ is taken to be independent of $t_0$. Thus, to lowest order we find
\begin{equation}
\label{L0rho0is0} \mathcal{L}_0 \rho_0 = 0.
\end{equation}
This implies $\rho_0$\ is a function of the conserved dynamical variables, notably the total
energy $H$, as well as of $R$.  The latter occurs because the derivatives $\displaystyle
\frac{\partial}{\partial r_i}$ and $\displaystyle \frac{\partial}{\partial p_i}$ in
$\mathcal{L}_0$ are to be taken at constant $R$, and thus $\mathcal{L}_0 R = 0$.  Then, we can
define $$ H = \sum_{i=1}^N \left ( \frac{p_i^2}{2m_i} +  V(\Gamma_r) \right )$$ and notice
that $\mathcal{L}_0 H = 0$.\\

Using the entropy maximization principle, one arrives at the nanocanonical solution to (\ref{L0rho0is0}) from \cite{MOa}:
\begin{equation}
\label{rho0} \rho_0 = \frac{e^{-\beta H}W(R,\underline{t})}{Q} \equiv \hat{\rho} W,
\end{equation}
where
\begin{equation}
\label{Q} Q(\beta, R) = \int  \Delta(\Gamma^\p_r, R) e^{-\beta H^\p} d \Gamma^\p.
\end{equation}
Here $\Delta$ is defined as in (\ref{delta}) and $H^\p$ is the $\Gamma^\p$-dependent value of
$H$.  For convenience we write $\underline{t}$ for the collection of slow time variables
$\underline{t} = \{t_1,t_2,...\}$.  As is standard in multiscale theory, determination of $W$ is
delayed until higher orders in the analysis.  With this, $\rho_0$ is seen to factorize into the
conditional probability $\hat{\rho}$ (i.e. for $\Gamma$ given $R$), multiplied by the reduced
probability $W$ for the slowly evolving state of the order parameter $R$.  We define the thermal
average of a given dynamical variable, $A(\Gamma)$ by
\begin{equation}
\label{thermal} A^{th} \equiv \int  \hat{\rho} \Delta \ A(\Gamma^\p) d \Gamma^\p.
\end{equation}
Now, we will assume that the nanocanonical ensemble obeys the Gibbs hypothesized equivalence
between the long-time and ensemble averages.  More specifically, we utilize a classical theorem
of Birkhoff (\cite{OW} can provide more detail) which states that the thermal average of a
dynamical variable $A(\Gamma)$ and its long-time average are equal.  Using classical semigroup
methods from applied partial differential equations (see \cite{RR} for more detail), one may
show that the linear operator $\mathcal{L}_0$ is the infinitesimal generator of a strongly
continuous semigroup on the function space $L^2(\Gamma)$.  This semigroup is then well-defined
and denoted by $\displaystyle e^{\mathcal{L}_0 t_0}$.  Hence, in the analysis that follows, we
will rely extensively on the property :
\begin{equation}
\label{GH} \lim_{t \rightarrow \infty} \frac{1}{t} \int_{-t}^0  e^{-\mathcal{L}_0 s} A ds =
A^{th}
\end{equation}
for all dynamical variables $A(\Gamma)$.  Thus, the long-time average or time evolution of a
variable does not affect the value of its thermal average as defined in (\ref{thermal}).  The
survey \cite{OW} or the classic article \cite{Hopf} can provide more background information and
detail from an ergodic theory perspective.\\

To $O(\epsilon)$ one finds

\begin{equation}
\label{II10}
\Lambda_0 \rho_1 = - \Lambda_1 \rho_0.
\end{equation}
Using the previously constructed semigroup $\displaystyle e^{\mathcal{L}_0 t_0}$, equation
(\ref{II10}) admits the solution
\begin{eqnarray*}
\label{II12} \rho_1 & = & e^{\mathcal{L}_0 t_0} \overline{A}_1 - \int_0^{t_0}
e^{\mathcal{L}_0(t_0 - t_0^\prime)} \Lambda_1 \rho_0 dt_0^\prime \\
& = & e^{\mathcal{L}_0 t_0} \overline{A}_1 - \int_0^{t_0}  e^{\mathcal{L}_0(t_0 - t_0^\prime)}
\left [\hat{\rho} \frac{\partial W}{\partial t_1} + \frac{P}{\hat{m}} \hat{\rho} \cdot
\frac{\partial W}{\partial R} + \frac{P}{\hat{m}} \cdot \frac{\partial \hat{\rho}}{\partial R} W
\right ] dt_0^\prime
\end{eqnarray*}
The first order initial condition $\overline{A}_1$ is, for now, undetermined and has the
dependence $\overline{A}_1(\Gamma,R,\underline{t})$.  As a consequence of the cross-level
communication inherent to multiscale analysis, the behavior of $\rho_1$ at large $t_0$ provides
information about the $t_1$-dependence of $W$, while the analysis of (\ref{consW}) provides a
necessary condition on $\overline{A}_1$ that ensures the equation determining $\wt{W}$ is
closed. Letting $s = t_0^\prime - t_0$, one obtains
\begin{equation}
\label{II13} \rho_1 = e^{\mathcal{L}_0 t_0} \overline{A}_1 - t_0 \hat{\rho} \frac{\partial
W}{\partial t_1} + \hat{\rho} \left [ \beta f^{th} W - \frac{\partial W}{\partial R} \right ]
\cdot \int_{-t_0}^0  e^{-\mathcal{L}_0 s} \frac{P}{\wh{m}} ds.
\end{equation}
In this equation, $\displaystyle \beta f^{th} = \frac{\partial}{\partial R} ( \ln Q )$, so that
$f^{th}$ is the force averaged via the nanocanonical ensemble.  This term occurs because of the
dependence of $\hat{\rho}$ on $R$ and we will verify the expression for $\displaystyle
\frac{\partial \hat{\rho}}{\partial R}$ in the Appendix. Thus, using the Gibbs Hypothesis
(\ref{GH}), we find
\begin{equation}
\label{II14} f^{th} = \lim_{t_0 \rightarrow \infty} \frac{1}{t_0} \int_{-t_0}^0
e^{-\mathcal{L}_0 s} f ds.
\end{equation}

Next, we remove secular behavior from the $t_0$-dependence in $\rho_1$, thereby imposing the
additional condition that $\rho_1$ remains bounded as $t_0 \rightarrow \infty$. Using
(\ref{II13}), it can be seen that if $\rho_1$ grows in $t_0$, it must do so at least linearly.
Hence, we may ensure that $\rho_1$ does not grow in $t_0$ by requiring that $\displaystyle
\lim_{t_0 \rightarrow \infty} \frac{1}{t_0} \rho_1 = 0$.  We then divide by $t_0$, take the
limit as $t_0 \rightarrow \infty$ in equation (\ref{II13}), and use (\ref{GH}). Notice that
$\displaystyle \left ( P \right )^{th} = 0$, as it involves terms of the form $\displaystyle
\int  p_i \ \exp \left (\frac{p_i^2}{2m_i} \right ) dp_i$.  Assuming the first order initial
data is taken in the nullspace of $\mathcal{L}_0$, that is $\mathcal{L}_0 \overline{A}_1 = 0$,
use of (\ref{GH}) yields $$\lim_{t_0 \rightarrow \infty} \frac{\rho_1}{t_0} = - \frac{\partial
W}{\partial t_1}.$$ Hence, we find
\begin{equation}
\label{Wt1}
\frac{\partial W}{\partial t_1} = 0
\end{equation}
and the reduced probability density $W$ is independent of $t_1$.  Note that this property follows
regardless of the choice of $\overline{A}_1$ in the nullspace of $\mathcal{L}_0$.  Using this in (\ref{II13}), we find

\begin{equation}
\label{rho1} \rho_1 = \overline{A}_1 + \hat{\rho} \left [ \beta f^{th} W - \frac{\partial
W}{\partial R} \right ] \int_{-t_0}^0 e^{-\mathcal{L}_0 s} \frac{P}{\wh{m}} ds
\end{equation}
concluding the $O(\epsilon)$ analysis, although $\overline{A}_1$ and $W$ are not yet
determined.\\

At this point, one would expect to conduct a $O(\epsilon^2)$ analysis of the problem and
determine an equation for $\displaystyle \frac{\partial W}{\partial t_2}$.  However, this is
unnecessary as the correction to the reduced probability density depends only on $\rho_0$ and
$\rho_1$ up to $O(\epsilon^2)$.  Instead, define for all $n=0,1,2,...$
\begin{equation}
\label{Wn} \wt{W}_n(R, t) = \int \Delta(\Gamma^\p_r,R) \rho_n(\Gamma^\p,t) d \Gamma^\p
\end{equation}
so that, using (\ref{Wtilde}) and (\ref{II4}), we may write
\begin{equation}
\label{Wexp} \wt{W} = \sum_{n=0}^\infty \epsilon^n \wt{W}_n.
\end{equation}
Hence, we expand $\wt{W}$ and $\rho$ in powers of $\epsilon$ as in (\ref{II4}) and
(\ref{Wexp}).  Using (\ref{rho0}) and (\ref{thermal}), the lowest order correction, $\wt{W}_0$, can be calculated as
\begin{eqnarray*}
\wt{W}_0 & = & \int \Delta(\Gamma^\p_r, R) \rho_0(\Gamma^\p,\underline{t}) d \Gamma^\p \\
& = & \int \Delta \rho_0(\Gamma^\p,\underline{t}, R^\p) d \Gamma^\p \\
& = & \int \Delta \hat{\rho} \ W(R^\p,\underline{t}) d \Gamma^\p \\
& = & W(R,\underline{t}).
\end{eqnarray*}

For $\epsilon \rightarrow 0$, one may see that $\wt{W} \rightarrow W$.  Hence, as the long time
scales tend to zero, the correction tends to the reduced probability density.  The $O(\epsilon)$
correction can be determined using (\ref{GH}) and (\ref{rho1}), so that

\begin{eqnarray*}
\wt{W}_1 & = & \int \Delta \rho_1(\Gamma^\p,t) d \Gamma^\p \\
& = & \int \Delta \left [ \overline{A}_1 - \hat{\rho} \left ( \frac{\partial
W}{\partial R} - \beta f^{th} W \right ) \cdot \int_{-t_0}^0 ds e^{-\mathcal{L}_0 s} \frac{P^\p}{\wh{m}} \right ] d \Gamma^\p \\
& = & \int \Delta \overline{A}_1(\Gamma^\p,R^\p,t) d \Gamma^\p \left. - \int \Delta \hat{\rho} \left ( \frac{\partial W}{\partial R}  - \beta f^{th} W \right ) \cdot \int_{-t_0}^0 e^{-\mathcal{L}_0 s} \frac{P^\p}{\wh{m}} ds d \Gamma^\p  \right \vert_{R = R^\p} \\
& = & \int \Delta \overline{A}_1(\Gamma^\p,R^\p,t) d \Gamma^\p + \left ( \int \Delta \hat{\rho} \int_{-t_0}^0 e^{-\mathcal{L}_0 s} \frac{P^\p}{\wh{m}} ds d \Gamma^\p \right ) \cdot \left ( - \frac{\partial W}{\partial R}(R,\underline{t})  + \beta f^{th} W(R,\underline{t}) \right ) \\
& = & \int \Delta \overline{A}_1(\Gamma^\p,R^\p,t) d \Gamma^\p + \left [\int_{-t_0}^0 \left
(e^{-\mathcal{L}_0 s} \frac{P}{\wh{m}}\right )^{th} ds \right ]
\cdot \left ( - \frac{\partial W}{\partial R}  + \beta f^{th} W \right )\\
& = & \int \Delta \overline{A}_1(\Gamma^\p,R^\p,t) d \Gamma^\p.
\end{eqnarray*}

Now, we may write $\rho(\Gamma,t)$ in terms of its expansion up to $O(\epsilon)$.  Using (\ref{rho0}) and (\ref{rho1}) in the right side of (\ref{consW}), we find
\begin{eqnarray*}
\int \Delta \frac{P^\p}{\wh{m}} \rho(\Gamma^\p,t) d\Gamma^\p & = & \int \Delta \frac{P^\p}{\wh{m}} \rho(\Gamma^\p,t,R^\p) d\Gamma^\p\\
& = & \int \Delta \frac{P^\p}{\wh{m}} \left ( \rho_0 + \epsilon \rho_1 \right ) d\Gamma^\p \\
& = & \int \Delta \hat{\rho} \frac{P^\p}{\wh{m}} W(R^\p,t) d\Gamma^\p +  \epsilon \int \Delta \frac{P^\p}{\wh{m}} \overline{A}_1(\Gamma^\p,R^\p,t) d\Gamma^\p \\
& \ & \ \ \ + \epsilon \left [ \int \Delta \hat{\rho} \frac{P^\p}{\wh{m}} \int_{-t_0}^0 e^{-\mathcal{L}_0 s} \frac{P^\p}{\wh{m}} ds \left ( \beta f^{th} W(R^\p,\underline{t}) - \frac{\partial W}{\partial R}(R^\p,\underline{t}) \right ) d\Gamma^\p \right ] \\
& = & \left ( \frac{P}{\wh{m}} \right )^{th} W(R,\underline{t}) +  \epsilon \int \Delta \frac{P^\p}{\wh{m}} \overline{A}_1(\Gamma^\p,R^\p,t) d\Gamma^\p \\
& \ & \ \ \ + \epsilon \left ( \frac{P^\p}{\wh{m}} \cdot \int_{-t_0}^0 e^{-\mathcal{L}_0 s} \frac{P^\p}{\wh{m}} ds \right )^{th} \left ( \beta f^{th} W(R,\underline{t}) - \frac{\partial W}{\partial R}(R,\underline{t}) \right ) \\
& = & \epsilon \int d\Gamma^\p \Delta \frac{P^\p}{\wh{m}} \overline{A}_1(\Gamma^\p,R^\p,t) +
\epsilon \frac{\gamma}{\hat{m}^2} \left ( \beta f^{th} W(R,\underline{t}) - \frac{\partial
W}{\partial R}(R,\underline{t}) \right )
\end{eqnarray*}
where the diffusion coefficient is
\begin{equation}
\label{gamma} \gamma = \int_{-t_0}^0 \left (P(0) \cdot P(s) \right )^{th} ds
\end{equation}
and we use the notation $P(s) = e^{-\mathcal{L}_0 s} P$.  Thus, (\ref{consW}) becomes
$$ \frac{\partial \wt{W}}{\partial t} = - \epsilon^2 \frac{\partial}{\partial R} \cdot \left ( \int \Delta \frac{P^\p}{\wh{m}} \overline{A}_1(\Gamma^\p,R^\p,t)  d\Gamma^\p \right )  + \epsilon^2 \frac{\partial}{\partial R} \cdot \left [\frac{\gamma}{\hat{m}^2} \left ( \frac{\partial W}{\partial R} - \beta f^{th} W \right ) \right ].$$  Using the expressions for $\wt{W}_0$ and $\wt{W}_1$, we can expand the reduced probability density as $\wt{W} = \wt{W}_0 + \epsilon \wt{W}_1$.  Then, isolating $\wt{W}_k$ terms and imposing the condition that $\overline{A}_1$ must stay bounded for large $R$, we find that this equation is closed only if $\overline{A}_1 = 0$.  Thus, up to $O(\epsilon^2)$, the conserved equation (\ref{consW}) becomes the Smoluchowski equation:
\begin{equation}
\label{Smoluchowski} \frac{\partial \wt{W}}{\partial t} =  \epsilon^2 \frac{\partial}{\partial
R} \cdot \left [\frac{\gamma}{\hat{m}^2} \left (\frac{\partial}{\partial R} - \beta f^{th}
\right ) \wt{W} \right ].
\end{equation}
Hence, in the case of fast-evolving momentum, due to $P$ and $f$ being $O(\epsilon^0)$, the
resulting behavior of the reduced probability density is governed by the Smoluchowski equation
(\ref{Smoluchowski}).  In the next section, we alter these assumptions on the behavior of
nanoparticle momentum and determine the corresponding changes in the structure of the equation
for $\wt{W}$.\\

\section{Multiscale Analysis: Fokker-Planck Equation}

In this section, we again use multiscale techniques to show that under similar circumstances,
the correction to the reduced probability density, $\wt{W}$, satisfies a Fokker-Planck equation.
In this situation, the momentum is not considered an atomistic variable, but instead as an order
parameter.  Hence, we define $\Gamma, m, R$, and $P$ as before, but reformulate the problem to
allow for the differing behavior of this slowly-evolving quantity.\\

To begin the multiscale analysis, we again introduce a dimensionless scaling parameter
$\epsilon$ in the mass terms by writing
\begin{equation}
\label{epsilonFP} \epsilon = \frac{\wh{m}}{m},
\end{equation}
where $\wh{m}$ is the mass of a typical atom. In this section, however, the assumptions on the
system of interest change.  We are now interested in significant migration distances on the
order of the nanoparticle diameter, which we take to be $O(\epsilon^{-\frac{1}{2}})$, and hence
scale $R$ to be $O(\epsilon^{-\frac{1}{2}})$.  Additionally, under the assumption that the
system is near equilibrium, the nanoparticle kinetic energy, $\displaystyle \frac{P^2}{2m}$ is
$O(\epsilon^0)$. Using the mass ratio scaling, this implies that $P =
O(\epsilon^{-\frac{1}{2}})$, as well. Finally, we assume that the net force on the nanoparticle
is reduced due to cancelation of atomic contributions, thus causing the momentum to evolve
slowly.  Hence, $f$ is assumed to be $O(\epsilon^\frac{1}{2})$.  A more detailed description
of these assumptions can be found in \cite{MOb}.\\

Under these considerations, define the scaled order parameters $\Phi$ and $\Pi$ by
\begin{equation}
\label{PhiFP}
\Phi = \epsilon^\frac{1}{2} R
\end{equation}
and
\begin{equation}
\label{PiFP}
\Pi = \epsilon^\frac{1}{2} P
\end{equation}
respectively.  The scaled net force $f$ can then be written in terms of the individual atomic
forces as
\begin{equation}
\label{fFP} f = \epsilon^{-\frac{1}{2}} \sum_{i=1}^N F_i \theta_i,
\end{equation}
and we let $V(\Gamma_r)$ be the $N$-atom potential so that $\displaystyle \frac{\partial V}{\partial
r_i} = -F_i$ for every $i = 1,...,N$.\\

Let us assume $\rho$ satisfies the Liouville Equation (\ref{Liouville}) where we again consider
$F_i$ to be the force on atom $i$ and $t$ to be time.  In addition, we denote the collection of
all atomic positions by $\Gamma_r = \{r_1,...,r_N\}$.  Given the probability density,
$\rho(\Gamma,t)$, we define
\begin{equation}
\label{WtildeFP}
\wt{W}(\Phi,\Pi,t) = \int \Delta(\Gamma^\p,\Phi,\Pi) \rho(\Gamma^\p,t) d\Gamma^\p .
\end{equation}
where
\begin{equation}
\label{deltaFP}
\Delta(\Gamma^\p,\Phi,\Pi) = \delta(\Phi - \Phi^\p) \delta(\Pi - \Pi^\p),
\end{equation} and the terms $$\Phi^\p = \epsilon^\frac{1}{2} \sum_{i=1}^N \frac{m_i}{m} r_i^\p \theta_i$$ and $$\Pi^\p = \epsilon^\frac{1}{2} \sum_{i=1}^N p_i^\p \theta_i$$ are the $\Gamma^\p$-dependent
values of $\Phi$ and $\Pi$.  Then, using the dependence of $\wt{W}$ on $\rho$, a solution to the Liouville
equation, we may show that $\wt{W}$ must satisfy a conserved equation similar to that of the previous section.  Since $\rho$ satisfies (\ref{Liouville}), we find
\begin{eqnarray*}
\frac{\partial \wt{W}}{\partial t} & = & \int \Delta \frac{\partial \rho}{\partial t}(\Gamma^\p,t) d \Gamma^\p \\
& = & - \int \Delta \left ( \sum_{i=1}^N \frac{p_i^\p}{m_i} \cdot \frac{\partial
\rho}{\partial r_i^\p} + F_i \cdot \frac{\partial \rho}{\partial
p_i^\p} \right ) d \Gamma^\p  \\
& = & \int \rho \left ( \sum_{i=1}^N \frac{p_i^\p}{m_i} \cdot \frac{\partial
\Delta}{\partial r_i^\p} + F_i \cdot \frac{\partial \Delta}{\partial p_i^\p}
\right ) d \Gamma^\p \\
& = & - \int \rho \left (\sum_{i=1}^N \frac{p_i^\p}{m_i} \cdot \frac{\partial
R^\p}{\partial r_i^\p} + F_i \cdot \frac{\partial R^\p}{\partial
p^\p_i} \right ) \epsilon^\frac{1}{2} \frac{\partial \Delta}{\partial \Phi} d \Gamma^\p \\
& \ & \ \ \ \ - \int \rho \left (\sum_{i=1}^N \frac{p_i^\p}{m_i} \cdot \frac{\partial P^\p}{\partial r_i^\p} + F_i \cdot \frac{\partial P^\p}{\partial p^\p_i} \right ) \epsilon^\frac{1}{2} \frac{\partial \Delta}{\partial \Pi} d \Gamma^\p \\
& = & \epsilon^\frac{1}{2} \frac{\partial}{\partial \Phi} \left ( \int \rho \Delta \mathcal{L}R^\p d \Gamma^\p \right ) + \epsilon^\frac{1}{2} \frac{\partial}{\partial \Pi} \left ( \int \rho \Delta \mathcal{L}P^\p d \Gamma^\p  \right )\\
& = & - \epsilon^\frac{1}{2} \frac{\partial}{\partial \Phi} \left ( \int \rho \Delta
\frac{P^\p}{m} d \Gamma^\p \right ) - \epsilon^\frac{1}{2} \frac{\partial}{\partial \Pi} \left ( \int \rho \Delta \sum_{i=1}^N F_i \theta_i d \Gamma^\p  \right )
\end{eqnarray*}
Thus, the reduced probability density, $\wt{W}$, satisfies
\begin{equation}
\label{consWFP} \frac{\partial \wt{W}}{\partial t} = - \epsilon \frac{\partial}{\partial \Phi}
\left ( \int \rho \Delta \frac{\Pi^\p}{\wh{m}} d \Gamma^\p  \right ) - \epsilon
\frac{\partial}{\partial \Pi} \left ( \int \rho \Delta f d \Gamma^\p  \right )
\end{equation}
which is in conservative form.\\

Next, we conduct a multiscale analysis in order to determine $\rho$ up to $O(\epsilon)$. Similar
to the previous section, the N-atom probability density, $\rho(\Gamma,t)$, is assumed to be
expressed as a function of two additional arguments, $\Upsilon(\Gamma,t,\cdot,\cdot)$ in such a
way that when the last two arguments are evaluated at $\Phi$ and $\Pi$, $\rho$ is obtained, i.e
$\rho(\Gamma,t) = \Upsilon(\Gamma,t,\Phi,\Pi)$.  Instead of labeling this new function, we will
just
extend our previous notation and refer to it as $\rho(\Gamma,t,\Phi,\Pi)$.\\

We apply the Liouville operator to $\rho(\Gamma,t,\Phi,\Pi)$ and invoke the chain rule to find
$$ \frac{\partial \rho}{\partial t} = - \mathcal{L}_0 \rho - \sum_{i=1}^N \frac{p_i}{m_i}
\cdot \frac{d\Phi}{d r_i} \frac{\partial \rho}{\partial \Phi} - \sum_{i=1}^N F_i \cdot
\frac{d\Pi}{d p_i} \frac{\partial \rho}{\partial \Pi}.$$ Using (\ref{PhiFP}) and (\ref{PiFP})
this becomes
\begin{equation}
\label{chainFP} \frac{\partial \rho}{\partial t} = - \mathcal{L}_0 \rho - \epsilon^\frac{1}{2}
\frac{P}{m} \frac{\partial \rho}{\partial \Phi} - \epsilon f \frac{\partial \rho}{\partial \Pi}.
\end{equation}
Here, we are writing $\mathcal{L}_0$ instead of $\mathcal{L}$ because these derivatives are
taken at constant values of $\Phi$ and $\Pi$.  By introducing (\ref{epsilonFP}) and (\ref{PiFP}) into (\ref{chainFP}), the Liouville equation (\ref{Liouville}) transforms into a multiscale equation as

\begin{equation}
\label{MultiscaleFP} \frac{\partial \rho}{\partial t} = \left ( \mathcal{L}_0 + \epsilon
\mathcal{L}_1 \right ) \rho
\end{equation}
where
\begin{equation}
\label{L0FP} \mathcal{L}_0 = - \sum_{i=1}^N \left [ \frac{p_i}{m_i} \cdot
\frac{\partial}{\partial r_i} + F_i \cdot \frac{\partial}{\partial p_i} \right ]
\end{equation}
and
\begin{equation}
\label{L1FP} \mathcal{L}_1 = - \frac{\Pi}{\wh{m}}\cdot \frac{\partial}{\partial \Phi} - f \cdot \frac{\partial}{\partial \Pi}.
\end{equation}
Again, it must be noted that $\mathcal{L}$ and $\mathcal{L}_0$, while seemingly exact in
definition, differ because the differentiation in $\mathcal{L}_0$ is performed at constant
values of order parameters $\Phi$ and $\Pi$.  Additionally, the differentiation in $\mathcal{L}_1$ is
performed at fixed values of $\Gamma$. Unlike the previous section, the operators (\ref{L0}) and
(\ref{L1}) are now the same as that of the previously mentioned papers \cite{MOa} and \cite{MOb}
since the conjugate momentum $\Pi$ is formulated as an order parameter.\\

Next, we assume that $\rho$ may be expressed as a power series in $\epsilon$:
\begin{equation}\label{II4FP}
\rho = \sum_{n=0}^\infty \rho_n \epsilon^n
\end{equation}
A set of time variables, defined via $t_n = \epsilon^n t$, is introduced to capture effects of
processes occurring on the various timescales. The chain rule implies
\begin{equation}\label{II5FP}
\frac{\partial}{\partial t} = \sum_{n=0}^\infty \epsilon^n \frac{\partial}{\partial t_n}.
\end{equation}
We then expand (\ref{MultiscaleFP}) using (\ref{II4FP}) and (\ref{II5FP}) and separate $\epsilon$
scales.  Define for $n \in \{0\} \cup \mathbb{N}$,
\begin{equation}
\label{LambdanFP} \Lambda_n = \frac{\partial}{\partial t_n} - \mathcal{L}_n
\end{equation}
where we take $\mathcal{L}_n = 0$ for $n > 1$.  The expansion yields the equations
$$ \Lambda_0 \rho_0 = 0,$$ and for $n \in \mathbb{N}$, $$ \Lambda_0 \rho_n = - \sum_{i=1}^n \Lambda_i
\rho_{n-i}.$$

Assuming the statistical state of the system has quasi-equilibrium character, the lowest order
distribution $\rho_0$ is taken to be independent of $t_0$. Thus, to lowest order we find
\begin{equation}
\label{L0rho0is0FP} \mathcal{L}_0 \rho_0 = 0.
\end{equation}
This implies $\rho_0$\ is a function of the conserved dynamical variables, notably the total
energy $H$, as well as of $\Phi$ and $\Pi$.  The latter occurs because the derivatives
$\displaystyle \frac{\partial}{\partial r_i}$ and $\displaystyle \frac{\partial}{\partial p_i}$
in $\mathcal{L}_0$ are to be taken at constant $\Phi$ and $\Pi$, and thus $\mathcal{L}_0 \Phi =
\mathcal{L}_0 \Pi = 0$.  As before, we define $$ H = \sum_{i=1}^N \left (
\frac{p_i^2}{2m_i} +  V(\Gamma_r) \right )$$ and notice that $\mathcal{L}_0 H = 0$.\\

Using the entropy maximization principle and proceeding as in \cite{MOa}, one arrives at the nanocanonical solution to (\ref{L0rho0is0FP}):
\begin{equation}
\label{rho0FP} \rho_0 = \frac{e^{-\beta H}W(\Phi,\Pi,\underline{t})}{Q} \equiv \hat{\rho} W,
\end{equation}
where
\begin{equation}
\label{QFP} Q(\beta, \Phi, \Pi) = \int \Delta(\Gamma^\p, \Phi, \Pi) e^{-\beta H^\p} d \Gamma^\p .
\end{equation}
This form of the nanocanonical solution is slightly different from that of \cite{MOa} since it was stated in that article that $Q$ is independent of $\Pi$.  We find that this is not the case and determine the exact manner in which the $\Pi$ dependence can be computed in the Appendix.  Here, $\Delta$ is defined as in (\ref{deltaFP}) and $H^\p$ is the $\Gamma^\p$-dependent value of $H$.  We label $\underline{t}$ as the collection of slow time variables $\underline{t} = \{t_1,t_2,...\}$.  As before, we define the thermal average of a given dynamical variable, $A(\Gamma)$ by
\begin{equation}
\label{thermalFP}
A^{th} = \int \Delta \hat{\rho} \ A(\Gamma^\p) d \Gamma^\p
\end{equation}
and use the Gibbs Hypothesis:
\begin{equation}
\label{GHFP} \lim_{t \rightarrow \infty} \frac{1}{t} \int_{-t}^0 e^{-\mathcal{L}_0 s} A ds =
A^{th}
\end{equation}
for all dynamical variables $A(\Gamma)$. Notice that the thermal average, and thus the
statements (\ref{thermalFP}) and (\ref{GHFP}), depend upon the new order parameter $\Pi$ because
of (\ref{deltaFP}).\\

To $O(\epsilon)$ one finds

\begin{equation}
\label{II10FP}
\Lambda_0 \rho_1 = - \Lambda_1 \rho_0
\end{equation}
and using the semigroup $e^{\mathcal{L}_0 t_0}$ defined in Section $2$, this equation admits the solution
\begin{eqnarray*}
\rho_1 & = & e^{\mathcal{L}_0 t_0} \overline{A}_1 - \int_0^{t_0} e^{\mathcal{L}_0(t_0
- t_0^\prime)} \Lambda_1 \rho_0 dt_0^\prime\\
& = & e^{\mathcal{L}_0 t_0} \overline{A}_1 - \int_0^{t_0} e^{\mathcal{L}_0(t_0 -
t_0^\prime)} \left [  \hat{\rho} \frac{\partial W}{\partial t_1} + \frac{\Pi}{\hat{m}}
\frac{\partial \hat{\rho}}{\partial \Phi} W + \frac{\Pi}{\hat{m}} \hat{\rho} \frac{\partial
W}{\partial \Phi} + f \frac{\partial \hat{\rho}}{\partial \Pi} W + f \hat{\rho} \frac{\partial
W}{\partial \Pi}\right ] dt_0^\prime \\
& = & e^{\mathcal{L}_0 t_0} \overline{A}_1 - \int_0^{t_0} e^{\mathcal{L}_0(t_0 -
t_0^\prime)} \left [  \hat{\rho} \frac{\partial W}{\partial t_1} - \frac{\Pi}{\hat{m}} \beta
\hat{\rho} f^{th} W + \frac{\Pi}{\hat{m}} \hat{\rho} \frac{\partial W}{\partial \Phi} + \beta f
\hat{\rho} \frac{\Pi}{\hat{m}} W + f \hat{\rho} \frac{\partial W}{\partial \Pi}\right ] dt_0^\prime
\end{eqnarray*}
where we have used the results $$\frac{\partial \hat{\rho}}{\partial \Phi} = - \beta \hat{\rho}
f^{th}$$ and $$\frac{\partial \hat{\rho}}{\partial \Pi} = \beta \hat{\rho}
\frac{\Pi}{\hat{m}}.$$ These derivatives will be verified in the Appendix.  The first order
initial condition $\overline{A}_1$ is, for now, undetermined and has the dependence
$\overline{A}_1(\Gamma,\Phi,\Pi,\underline{t})$. As a consequence of the cross-level
communication inherent to multiscale analysis, the behavior of $\rho_1$ at large $t_0$ provides
information about the $t_1$-dependence of $W$, while the analysis of (\ref{consWFP}) provides a
necessary condition on $\overline{A}_1$ that ensures the equation determining $\wt{W}$ is
closed. Letting $s = t_0^\prime - t_0$, one obtains

\begin{eqnarray}
\rho_1 & = & e^{\mathcal{L}_0 t_0} \overline{A}_1 - t_0 \hat{\rho} \frac{\partial W}{\partial
t_1} - \hat{\rho} \beta \frac{\Pi}{\hat{m}} W \cdot \int_{-t_0}^0 e^{-\mathcal{L}_0 s}
\left ( f - f^{th} \right ) ds \notag \\
\label{II13FP} & \ &  \ \ \  - t_0 \hat{\rho} \frac{\Pi}{\hat{m}} \cdot \frac{\partial
W}{\partial \Phi} - \hat{\rho} \frac{\partial W}{\partial \Pi} \cdot \int_{-t_0}^0
e^{-\mathcal{L}_0 s} f ds.
\end{eqnarray}

Next, we remove secular behavior from the $t_0$-dependence in $\rho_1$.  As before, we assume
that $\rho_1$ remains bounded as $t_0 \rightarrow \infty$. Using (\ref{II13FP}), it can be seen
that if $\rho_1$ grows in $t_0$, it must do so at least linearly.  Hence, we may ensure that
$\rho_1$ does not grow in $t_0$ by requiring that $\displaystyle \lim_{t_0 \rightarrow \infty}
\frac{1}{t_0} \rho_1 = 0$.  We then divide by $t_0$, take the limit as $t_0 \rightarrow \infty$
in the above equation, and use (\ref{GHFP}). Assuming the first order initial data is in the
nullspace of $\mathcal{L}_0$, this yields
$$\lim_{t_0 \rightarrow \infty} \frac{\rho_1}{t_0} = - \frac{\partial W}{\partial t_1} - \frac{\Pi}{\hat{m}} \cdot \frac{\partial W}{\partial \Phi} - f^{th} \cdot \frac{\partial W}{\partial \Pi}.$$ Hence, we find
\begin{equation}
\label{Wt1FP} \Lambda_1^{th} \equiv \frac{\partial W}{\partial t_1} + \frac{\Pi}{\hat{m}} \cdot
\frac{\partial W}{\partial \Phi} + f^{th} \cdot \frac{\partial W}{\partial \Pi} = 0
\end{equation}
and $W$ satisfies a Liouville equation in $(t_1,\Phi,\Pi)$ space.  Note that (\ref{Wt1FP})
follows regardless of the choice of $\overline{A}_1$ in the nullspace of $\mathcal{L}_0$.  Using
this in (\ref{II13FP}), we find

\begin{equation}
\label{rho1FP} \rho_1 = \overline{A}_1 - \hat{\rho} \left [ \beta \frac{\Pi}{\hat{m}} W +
\frac{\partial W}{\partial \Pi} \right ] \int_{-t_0}^0 e^{-\mathcal{L}_0 s} \left ( f -
f^{th} \right ) ds
\end{equation}
concluding the $O(\epsilon)$ analysis, although $\overline{A}_1$ and $W$ are not yet determined.\\

As before, the correction to the reduced probability density depends only on $\rho_0$ and
$\rho_1$ up to $O(\epsilon^2)$.  Hence, define for every $n=0,1,2,...$
\begin{equation}
\label{WnFP}
\wt{W}_n(\Phi,\Pi,t) = \int \Delta(\Gamma^\p,\Phi,\Pi) \rho_n(\Gamma^\p,t) d \Gamma^\p
\end{equation}
so that, using (\ref{WtildeFP}) and (\ref{II4FP}), we may write
\begin{equation}
\label{WexpFP} \wt{W} = \sum_{n=0}^\infty \epsilon^n \wt{W}_n.
\end{equation}
In addition, we may expand $\wt{W}$ and $\rho$ in powers of $\epsilon$ as in (\ref{II4FP}) and
(\ref{WexpFP}).  Using (\ref{GHFP}), the lowest order correction, $\wt{W}_0$, can be calculated as
\begin{eqnarray*}
\wt{W}_0 & = & \int \Delta(\Gamma^\p, \Phi, \Pi) \rho_0(\Gamma^\p,\underline{t}) d \Gamma^\p \\
& = & \int \Delta \rho_0(\Gamma^\p,\underline{t}, \Phi^\p, \Pi^\p) d \Gamma^\p \\
& = & \int  \Delta \hat{\rho} \ W(\Phi^\p, \Pi^\p,\underline{t}) d \Gamma^\p \\
& = & W(\Phi,\Pi,\underline{t}).
\end{eqnarray*}

For $\epsilon \rightarrow 0$, one may see that $\wt{W} \rightarrow W$.  Hence, as the long time
scales tend to zero, the correction tends to the reduced probability density.  The $O(\epsilon)$
correction can be determined using (\ref{thermalFP}).  Notice that (\ref{GHFP}) implies $\left ( e^{-\mathcal{L}_0 \tau} A \right )^{th} = A^{th}$ for any finite value of $\tau \in \mathbb{R}$. Using this and (\ref{rho1FP}), we find

\begin{eqnarray*}
\wt{W}_1 & = & \int \Delta \rho_1(\Gamma^\p,t) d \Gamma^\p \\
& = & \int \Delta \left [ \overline{A}_1 - \hat{\rho} \left ( \frac{\partial
W}{\partial \Pi}  + \beta \frac{\Pi^\p}{\hat{m}} W \right ) \cdot \int_{-t_0}^0 ds e^{-\mathcal{L}_0
s} \left ( f - f^{th} \right )\right ] d \Gamma^\p \\
& = & \int \Delta \overline{A}_1(\Gamma^\p,\Phi^\p,\Pi^\p, t) d \Gamma^\p \\
& \ & \ \ \ \left. - \int \Delta \hat{\rho} \left ( \frac{\partial W}{\partial \Pi}
+ \beta \frac{\Pi^\p}{\hat{m}} W \right ) \cdot \int_{-t_0}^0 e^{-\mathcal{L}_0 s} \left (f -
f^{th} \right ) ds d \Gamma^\p \right \vert_{(\Phi,\Pi) = (\Phi^\p, \Pi^\p)} \\
& = & \int \Delta \overline{A}_1(\Gamma^\p,\Phi^\p,\Pi^\p,t) d \Gamma^\p - \left [ \int \Delta
\hat{\rho} \int_{-t_0}^0 e^{-\mathcal{L}_0 s} \left (f - f^{th} \right ) ds d \Gamma^\p
 \right ] \cdot \left ( \frac{\partial W}{\partial \Pi}(\Phi,t)  + \beta \frac{\Pi}{\hat{m}} W(\Phi,t) \right ) \\
& = & \int \Delta \overline{A}_1(\Gamma^\p,\Phi^\p,\Pi^\p,t) d \Gamma^\p - \int_{-t_0}^0 \left (f(s) -
f^{th} \right )^{th} ds \cdot
\left ( \frac{\partial W}{\partial \Pi}  + \beta \frac{\Pi}{\hat{m}} W \right )\\
& = & \int \Delta \overline{A}_1(\Gamma^\p,\Phi^\p, \Pi^\p,t)  d \Gamma^\p
\end{eqnarray*}
where we use the notation $f(s) = e^{-\mathcal{L}_0 s} f$. Now, we may write
$\rho(\Gamma,t)$ in terms of its expansion up to $O(\epsilon)$.  Using (\ref{rho0FP}) and
(\ref{rho1FP}) in the first term on the right side of (\ref{consWFP}), we find
\begin{eqnarray*}
\int \Delta \frac{\Pi^\p}{\wh{m}} \rho(\Gamma^\p,t) d\Gamma^\p & = & \int \Delta
\frac{\Pi^\p}{\wh{m}} \rho(\Gamma^\p,t,\Phi^\p,\Pi^\p) d\Gamma^\p \\
& = & \int \Delta \frac{\Pi^\p}{\wh{m}} \left ( \rho_0 + \epsilon \rho_1 \right ) d\Gamma^\p \\
& = & \int \Delta \hat{\rho} \frac{\Pi^\p}{\wh{m}} W(\Phi^\p,\Pi^\p,t) d\Gamma^\p + \epsilon
\int \Delta \frac{\Pi^\p}{\wh{m}} \overline{A}_1(\Gamma^\p,\Phi^\p,\Pi^\p,t) d\Gamma^\p \\
& \ & \ \ \ - \epsilon \left [ \int \Delta \hat{\rho} \frac{\Pi^\p}{\wh{m}} \left (
\beta \frac{\Pi^\p}{\hat{m}} W + \frac{\partial W}{\partial \Pi} \right ) \cdot \int_{-t_0}^0 ds
e^{-\mathcal{L}_0 s} \left ( f - f^{th} \right ) d\Gamma^\p \right ] \\
& = & \frac{\Pi}{\wh{m}} W(\Phi,\Pi,t) +  \epsilon \frac{\Pi}{\wh{m}} \int \Delta
\overline{A}_1(\Gamma^\p,\Phi^\p,\Pi^\p,t) d\Gamma^\p \\
& \ & \ \ \ - \epsilon \frac{\Pi}{\wh{m}} \left ( \beta \frac{\Pi}{\hat{m}} W + \frac{\partial
W}{\partial \Pi} \right ) \cdot \left ( \int_{-t_0}^0 (f(s) - f^{th} ) ds \right )^{th}\\
& = & \frac{\Pi}{\wh{m}} W(\Phi,\Pi,t) +  \epsilon \frac{\Pi}{\wh{m}} \int \Delta
\overline{A}_1(\Gamma^\p,\Phi^\p,\Pi^\p,t) d\Gamma^\p.
\end{eqnarray*}
Similarly, the second term on the right side of (\ref{consW}) becomes
\begin{eqnarray*}
\int \Delta f \rho(\Gamma^\p,t) d \Gamma^\p & = & \int \Delta
f \rho(\Gamma^\p,t,\Phi^\p,\Pi^\p) d\Gamma^\p \\
& = & \int \Delta f \left ( \rho_0 + \epsilon \rho_1 \right ) d\Gamma^\p \\
& = & \int \Delta \hat{\rho} f  W(\Phi^\p,\Pi^\p,\underline{t}) d\Gamma^\p  +  \epsilon \int
\Delta f \overline{A}_1(\Gamma^\p,\Phi^\p,\Pi^\p,t) d\Gamma^\p\\
& \ & \ \ \ - \epsilon \left [ \int \Delta \hat{\rho} f \left ( \beta
\frac{\Pi^\p}{\hat{m}} W + \frac{\partial W}{\partial \Pi} \right ) \cdot \int_{-t_0}^0 ds
e^{-\mathcal{L}_0 s} \left ( f - f^{th} \right ) d\Gamma^\p \right ] \\
& = & f^{th} W(\Phi,\Pi,\underline{t}) +  \epsilon \int \Delta
f \overline{A}_1(\Gamma^\p,\Phi^\p,\Pi^\p,t) d\Gamma^\p \\
& \ & \ \ \ - \epsilon \left ( \beta \frac{\Pi}{\hat{m}} W + \frac{\partial W}{\partial \Pi}
\right ) \left ( \int_{-t_0}^0 f(0) \cdot (f(s) - f^{th} ) ds \right )^{th}\\
& = & f^{th} W(\Phi,\Pi,\underline{t}) +  \epsilon \int \Delta f \overline{A}_1(\Gamma^\p,\Phi^\p,\Pi^\p,t)  d\Gamma^\p - \epsilon \theta \left ( \beta \frac{\Pi}{\hat{m}} W + \frac{\partial W}{\partial \Pi} \right )
\end{eqnarray*}
where
\begin{equation}
\label{theta} \theta = \int_{-t_0}^0 \left [ ( f(0) \cdot f(s))^{th} - f^{th} \cdot f^{th} \right ] ds.
\end{equation}
Thus, (\ref{consWFP}) becomes
\begin{eqnarray*}
\frac{\partial \wt{W}}{\partial t} & = & - \epsilon \frac{\partial}{\partial \Phi} \cdot \left (
\frac{\Pi}{\wh{m}} W(\Phi,\Pi,\underline{t}) \right )  - \epsilon^2 \frac{\partial}{\partial \Phi} \cdot \left ( \frac{\Pi}{\wh{m}} \int \Delta \overline{A}_1(\Gamma^\p,\Phi^\p,\Pi^\p,t) d\Gamma^\p \right ) \\
& \ & \ \ \ - \epsilon \frac{\partial}{\partial \Pi} \cdot \left ( f^{th} W(\Phi,\Pi,\underline{t}) \right ) +
\epsilon^2 \frac{\partial}{\partial \Pi} \cdot \left [ - \int \Delta f
\overline{A}_1(\Gamma^\p,\Phi^\p,\Pi^\p,t) d\Gamma^\p  + \theta \left ( \beta \frac{\Pi}{\hat{m}} W + \frac{\partial W}{\partial \Pi} \right )\right ].
\end{eqnarray*}
Using the expressions for $\wt{W}_0$ and $\wt{W}_1$, we can expand the reduced probability density as $\wt{W} = \wt{W}_0 + \epsilon \wt{W}_1$.  Then, isolating $\wt{W}_k$ terms and imposing the condition that $\overline{A}_1$ must stay bounded for large $\Phi$ and $\Pi$, we find that this equation is closed only if $\overline{A}_1 = 0$.  Thus, up to $O(\epsilon^2)$, the conserved equation (\ref{consWFP}) becomes the Fokker-Planck equation:

\begin{equation}
\label{FP} \frac{\partial \wt{W}}{\partial t} =  - \epsilon \frac{\partial}{\partial \Phi} \cdot \left (
\frac{\Pi}{\wh{m}} \wt{W} \right ) - \epsilon \frac{\partial}{\partial \Pi} \cdot \left ( f^{th} \wt{W} \right ) + \epsilon^2 \frac{\partial}{\partial \Pi} \cdot \left [ \theta \left ( \beta \frac{\Pi}{\hat{m}}  + \frac{\partial}{\partial \Pi} \right )\wt{W} \right ].
\end{equation}
Hence, under the assumption that momenta have (comparatively) large values and evolve slowly,
i.e., $P = O(\epsilon^{-\frac{1}{2}})$ and  $f = O(\epsilon^\frac{1}{2})$, the reduced
probability
density obeys a Fokker-Planck equation given by (\ref{FP}).\\

\section{Appendix}

We first verify the $\Phi$ derivative of $\hat{\rho}$ used in the derivation of both equations.
In Section $2$, the variable $R$ is used instead of $\Phi$ (as order parameters are unscaled
with respect to $\epsilon$), but the statements that follow can be applied exactly to $R$ in the
same manner as $\Phi$. We claim
$$ \frac{\partial \hat{\rho}}{\partial \Phi} = -\beta \hat{\rho} f^{th}.$$ Using (\ref{rho0}) or
(\ref{rho0FP}), we see that
$$ \frac{\partial \hat{\rho}}{\partial \Phi} =  - \frac{e^{-\beta H}}{Q^2} \ \frac{\partial Q}{\partial \Phi}.$$  Notice that $\displaystyle \frac{\partial}{\partial \Phi} \left ( e^{-\beta H} \right ) = 0$ since $\Phi$ derivatives are to be taken at constant values of $\Gamma$.  Using (\ref{Q}) or (\ref{QFP}), we see that
\begin{eqnarray*}
\frac{\partial Q}{\partial \Phi} & = & \int \frac{\partial \Delta}{\partial \Phi} e^{-\beta H^\p} d\Gamma^\p \\
& = & - \int \frac{\partial \Delta}{\partial \Phi^\p} e^{-\beta H^\p} d\Gamma^\p \\
& = & \epsilon^{-\frac{1}{2}} \int \Delta \frac{\partial}{\partial R^\p} \left (e^{-\beta H^\p} \right ) d\Gamma^\p
\end{eqnarray*}
Now, since this integration is performed over all values of $\Gamma^\p$, and hence $\Gamma^\p$
is not fixed, the energy depends upon the center of mass through the potential function. Notice,
we may calculate $\displaystyle \frac{\partial V}{\partial R}$ as $V$ depends upon $R$
implicitly in the following manner.  For every $i=1,...,N$, write the residual displacement of
each atomic position as $r_i = \sigma_i  + R \theta_i$.  Here the $N$ variables $\{R,
\sigma_1,..., \sigma_{N-1}\}$ constitute a complete set of variables as we may write $\sigma_N$
in terms of each of the other $\sigma_i$ using the constraint
$$\sum_{i=1}^N \sigma_i \theta_i = 0,$$ which follows by the definition of the residual
coordinates.  This change of variables is just a set of linear functions with constant
coefficients, hence the Jacobian is constant and $$ \frac{\partial V}{\partial R} = \sum_{i=1}^N
\frac{\partial V}{\partial r_i} \theta_i = - \epsilon^\frac{1}{2} f.$$  Since the kinetic energy
is independent of $\Gamma_r$, we find
$$ \frac{\partial}{\partial R^\p} \left (e^{-\beta H^\p} \right ) = -\beta e^{-\beta H^\p}
\frac{\partial V^\p}{\partial R^\p} = \beta \epsilon^\frac{1}{2} f^\p e^{-\beta H^\p}.$$  Using
this, the derivative of $Q$ becomes
\begin{eqnarray*}
\frac{\partial Q}{\partial \Phi} & = & \epsilon^{-\frac{1}{2}} \int \Delta e^{-\beta H^\p} \beta \epsilon^\frac{1}{2} f^\p d\Gamma^\p \\
& = & \beta \int \Delta e^{-\beta H^\p} f^\p d\Gamma^\p.
\end{eqnarray*}
Hence, using the modified Gibbs Hypotheses in either section (\ref{GH}) or (\ref{GHFP}), we find
\begin{eqnarray*}
\frac{\partial \hat{\rho}}{\partial \Phi} & = & - \frac{e^{-\beta H}}{Q^2} \ \frac{\partial Q}{\partial \Phi} \\
& = & - \hat{\rho} \left ( \frac{1}{Q} \cdot \frac{\partial Q}{\partial \Phi} \right ) \\
& = & - \hat{\rho} \beta \int \frac{1}{Q} \Delta e^{-\beta H^\p} f^\p d\Gamma^\p \\
& = & - \beta \hat{\rho} \left ( \int \Delta \hat{\rho} f^\p d\Gamma^\p \right ) \\
& = & - \beta \hat{\rho} f^{th}
\end{eqnarray*}
and the verification of this derivative is complete.\\

Next, we verify the $\Pi$ derivative of $\hat{\rho}$ used in Section $3$.  We claim
$$ \frac{\partial \hat{\rho}}{\partial \Pi} = \beta \hat{\rho} \frac{\Pi}{\hat{m}}.$$
First, we use (\ref{QFP}) and rewrite $Q$ in terms of its $\Pi$ dependence using the Fourier
transform. Since the potential and kinetic energies depend exclusively on $r_i$ and $p_i$
variables respectively, we separate $Q$ into two different integrals involving these variables.
In addition, we separate the momentum integrals into the momenta of particles in the host
medium, denoted $\displaystyle \Gamma_0 = \{ p_i : \theta_i = 0 \}$ and those in the
nanoparticle, denoted $\displaystyle \Gamma_1 = \{ p_i : \theta_i = 1 \}$.  Using the notation
$$ K_0 = \sum_{i=1}^N \frac{p_i^2}{2m_i} (1 - \theta_i)$$ and $$ K_1 =
\sum_{i=1}^N \frac{p_i^2}{2m_i} \theta_i $$ for the kinetic energies of the host medium and nanoparticle, respectively, we write $Q$ as
\begin{equation}
\label{separateQ}
Q(\beta,\Phi,\Pi) = \left ( \int \delta(\Phi - \Phi^\p) e^{-\beta V^\p} d\Gamma^\p_r \right ) \left ( \int e^{-\beta K_0^\p} d\Gamma^\p_0 \right ) \left ( \int \delta(\Pi - \Pi^\p) e^{-\beta K_1^\p} d\Gamma^\p_1 \right ).
\end{equation}
Keeping the $\Gamma_r^\p$ and $\Gamma_0^\p$ integrals as they are (notice further that the $\Gamma_0^\p$ integral is constant), we focus on the $\Gamma_1^\p$ integral.  We may again split the $\Gamma_1^\p$ integral into one each in the $x,y$, and $z$ directions.  We will consider the integral in the $x$-direction, labeled $I_x$, and state that the results we obtain will follow for the integrals in the other directions in the same manner.\\

Now, we relabel the momenta in the nanoparticle, $1$ through $M$, and write their $x$-coordinates as $x_1$ through $x_M$. Then, write the $x$-directional Dirac mass $\delta(\Pi_x - \Pi_x^\p)$ using the Fourier transform of the function $f(k) = 1$ as $$\delta (\Pi_x - \Pi_x^\p) = \frac{1}{\sqrt{2\pi}} \int_{-\infty}^\infty \exp \left [\displaystyle ik \left ( \Pi_x - \sqrt{\epsilon} \sum_{l=1}^M x_l \right ) \right ] \ dk.$$
Then, using the Inverse Fourier Transform on the resulting Gaussians
$$ \exp(-\lambda k^2) = \frac{1}{\sqrt{2\pi}} \int_{-\infty}^\infty \frac{1}{\sqrt{4\pi \lambda}} \exp \left ( -ikz  \right ) \exp \left ( -\frac{z^2}{4\lambda} \right ) \ dz.$$
and (\ref{epsilonFP}), we find
\begin{eqnarray*}
I_x & = & \frac{1}{\sqrt{2\pi}} \int \int_{-\infty}^\infty \exp \left [\displaystyle ik \left ( \Pi_x - \sqrt{\epsilon} \sum_{l=1}^M x_l \right ) \right ] \ dk \\
& \ & \ \ \ \cdot \exp \left ( -\frac{\beta}{2m_1} x_1^2 \right ) \cdot \cdot \cdot \exp \left ( -\frac{\beta}{2m_M} x_M^2 \right ) dx_1\cdot\cdot\cdot dx_M\\
& = & \frac{1}{\sqrt{2\pi}} \int_{-\infty}^\infty \exp \left (ik \Pi_x \right )   \left [ \int \exp \left (-ik \sqrt{\epsilon} x_1 \right ) \exp \left (-\frac{\beta}{2m_1} x_1^2\right ) d x_1 \right ] \cdot \cdot \cdot \\
& \ & \ \ \ \cdot \cdot \cdot \left [ \int \exp \left (-ik \sqrt{\epsilon} x_M \right ) \exp \left (-\frac{\beta}{2m_M} x_M^2 \right ) d x_M \right ] dk\\
& = & \int_{-\infty}^\infty e^{ik \Pi_x} \left [ \sqrt{\frac{m_1}{\beta}} \ e^{\displaystyle -\frac{\epsilon m_1 k^2}{2\beta}} \right ] \cdot\cdot\cdot \left [ \sqrt{\frac{m_M}{\beta}} \ e^{\displaystyle -\frac{\epsilon m_M k^2}{2\beta}} \right ] dk \\
& = & \sqrt{\frac{m_1}{2\pi\beta}} \cdot\cdot\cdot \sqrt{\frac{m_M}{2\pi\beta}} \int_{-\infty}^\infty \displaystyle \exp \left ( ik \Pi_x \right ) \exp \left [-\frac{\epsilon k^2}{2\beta} m \right ] dk \\
& = & C_1 \exp \left ( -\frac{\beta}{2\hat{m}} \Pi_x^2 \right )
\end{eqnarray*}
where $$C_1 = \sqrt{\frac{m_1}{2\pi\beta}} \cdot\cdot\cdot \sqrt{\frac{m_M}{2\pi\beta}} \cdot \sqrt{\frac{2\pi\beta}{\hat{m}}}.$$
We extend this in the $y$ and $z$ directions and multiply to find
$$ \int \delta(\Pi - \Pi^\p) e^{-\beta K_1^\p} d\Gamma^\p_1  = (C_1)^3 \exp \left ( -\frac{\beta}{2\hat{m}} \Pi^2 \right ).$$ Thus, we can write
$$ Q(\beta,\Phi,\Pi) = Q_1(\Phi) (C_1)^3 C_2 \exp \left ( -\frac{\beta}{2\hat{m}} \Pi^2 \right ) $$ where $$ Q_1(\Phi) = \left ( \int  \delta(\Phi - \Phi^\p) e^{-\beta V^\p} d\Gamma^\p_r \right )$$ and  $$C_2 = \left ( \int e^{-\beta K_0^\p} d\Gamma^\p_0 \right ).$$  Finally, $\hat{\rho}$ can be expressed in the form
\begin{equation}
\label{rhohatalt}
\hat{\rho} = \frac{e^{-\beta H}}{Q_1(\Phi) (C_1)^3 C_2}
\exp \left ( \frac{\beta}{2\hat{m}} \Pi^2 \right ).
\end{equation}
Taking a $\Pi$ derivative in (\ref{rhohatalt}), which must be done at fixed values of $\Gamma$, we find $$ \frac{\partial \hat{\rho}}{\partial \Pi} = \beta \hat{\rho} \frac{\Pi}{\hat{m}}$$ and the verification of this derivative is complete.

\end{document}